# Finite-size effects and energy alignment in molecular XANES under periodic boundary conditions: A systematic comparison of core-hole treatments

Yu Fujikata[1,2,a)], Yasuji Muramatsu[3] and Teruyasu Mizoguchi[1,a)]

[1]Institute of Industrial Science, The University of Tokyo, Tokyo 153-8505, Japan.
[2]Science & Innovation Center, Mitsubishi Chemical Corporation, Yokohama 227-8502, Japan.
[3]Graduate School of Engineering and Laboratory of Advanced Science and Technology for Industry (LASTI), University of Hyogo 2146 Shosha, Himeji 671–2201, Japan.

**Abstract**

X-ray absorption near-edge structure (XANES) provides element-specific insight into local electronic and structural environments, but quantitative interpretation of molecular XANES under periodic boundary conditions (PBC) remains challenging due to finite-size effects and core-hole treatments. In this work, we systematically investigate how core-hole approximations and charge compensation schemes affect transition energies, energy alignment, and chemical-shift reproducibility in PBC-DFT-based molecular XANES calculations. Using ethane as a model system, we show that the full core-hole (FCH) approach exhibits pronounced supercell-size dependence originating from interactions between background charge and charged molecules, with transition energies largely changed by leading-order finite-size terms. In contrast, the excited core-hole (XCH) method rapidly converges owing to its neutral final state. We further demonstrate that most finite-size effects in FCH can be removed by Makov–Payne corrections based on multipole expansion of the electrostatic energy of charged supercells under PBC. Furthermore, we propose a simple Fermi-level-based energy correction ($E_F/2$) that provides comparable improvement using only a single supercell. Extending the analysis to an *n*-alkane series reveals that while intrinsic electronic-structure changes govern peak shifts for small molecules, systematic energy drifts persist in FCH for larger molecules, whereas XCH and FCH+$E_F/2$ remain stable. Finally, for small molecules at the C and N K-edges, XCH and FCH+$E_F/2$ accurately reproduce experimental chemical shifts, whereas uncorrected FCH fails. These results provide practical guidelines for reliable energy alignment and chemical-shift analysis in molecular XANES under PBC, supporting robust applications to molecular, adsorption, and interfacial systems.

**Introduction**

X-ray absorption spectroscopy (XAS) enables simultaneous characterization of local atomic structure and electronic states with elemental specificity, and has therefore been widely applied to the analysis of functional materials including batteries, catalysts, and ceramics.[1–4] In particular, X-ray absorption near-edge structure (XANES) is highly sensitive to the oxidation state, coordination environment, local symmetry, and unoccupied electronic states of the absorbing atom, and has long served as an essential probe, especially for inorganic materials.[5] Recent advances in synchrotron radiation sources, detectors, and measurement methodologies have significantly expanded the applicability of soft X-ray XANES, extending its use to organic molecules, organic materials, and interfacial or adsorption systems. For example, C, N, and O K-edge XANES has been employed to identify surface functional groups formed on carbon fibers during electrochemical treatment.[6] In addition, angle-dependent C K-edge XANES measurements have been used to evaluate molecular orientation in two-dimensional thin films of p-type organic semiconductors utilized in organic light-emitting diode devices.[7] Furthermore, taking advantage of its high experimental flexibility, operando XANES measurements under catalytic reaction conditions have become increasingly common; notably, carbonate anion adsorption on Co–$O_x$ water-splitting catalysts has been directly observed.[8]

Despite its high sensitivity to subtle structural and electronic variations, the XANES spectral line shape, including peak positions and intensities, is governed by a complex interplay of many-body effects, core-hole screening, final-state interactions, and the distribution of delocalized unoccupied states. Consequently, direct and unambiguous interpretation of experimental XANES spectra based solely on intuition remains challenging, and reliable analysis generally requires comparison with theoretical calculations.[9]

Several theoretical approaches have been developed for XANES calculations, including methods based on multiple-scattering theory,[10,11] approaches that evaluate transition probabilities using molecular orbital calculations,[12,13] and band-structure-based methods under periodic boundary conditions (PBC).[14,15] For molecular and cluster systems, non-periodic frameworks such as multiple-scattering theory and molecular orbital methods have traditionally been widely employed.[16] In recent years, however, the rapid growth of computational resources, advances in parallel computing, and acceleration using GPUs have enabled the placement of molecules or clusters within large supercells and their treatment using density functional theory (DFT) band-structure codes. In other words, non-periodic structures can now be described within three-dimensional periodic boundary conditions in practical computational timescales.[17] This development has broadened the scope of PBC-based XANES simulations beyond bulk crystals. This capability is particularly important for XANES applications, which, as mentioned above, are often performed under operando surface conditions. Realistic simulations of XANES for adsorption systems, where non-periodic molecules are placed on crystalline surfaces with two-dimensional periodicity, have therefore become feasible. A major advantage of the band-structure framework is that it enables a unified treatment of 3D periodic crystals, 2D periodic surfaces and interfaces, and non-periodic molecules, which is particularly useful for catalysis- and interface-related systems. Motivated by this background, recent studies have reported large-scale calculations and analyses of XANES spectra for extensive molecular datasets using band-structure-based approaches.[18]

While advances in computational methodologies have substantially improved the reproduction of experimental spectral line shapes, accurate treatment of transition energies remains essential when comparing theoretical and experimental XANES spectra. Although first-principles calculations generally struggle to reproduce absolute transition energies, reliable prediction of relative shifts, namely chemical shifts, is often sufficient for practical interpretation. A representative theoretical framework for evaluating transition energies is the ΔSCF approach, in which excitation energies are obtained from total energy differences between the ground state and the excited state, where a core electron is promoted or removed.[19] For describing the excited state, several schemes have been proposed, including the full core-hole (FCH) method, which explicitly removes a core electron,[20–22] and the excited core-hole (XCH) approach, in which the excited electron is placed at the bottom of the conduction band, as schematically shown in Figure 1.[23,24]

When the FCH approximation is employed under periodic boundary conditions, removal of a core electron renders the unit cell positively charged, leading to a divergence of the Coulomb energy. To stabilize such calculations, it is common practice to introduce a uniform compensating background charge, often referred to as jellium.[25–27] However, this background charge interacts with the localized positive charge associated with the core hole, thereby introducing finite-size effects into the total energy of the system. As a consequence, the calculated transition energies, namely the energy shifts, can exhibit an artificial dependence on the supercell size. In contrast, within the XCH framework, the excited electron is retained within the system, allowing construction of an excited state that remains globally charge-neutral. This eliminates the need for a compensating background charge and constitutes a key advantage of the XCH approach. Therefore, even for the same XANES calculation, the choice of core-hole approximation and the treatment of charge compensation in periodic systems can give rise to systematic differences in the simulated transition energies.

It should be noted that finite-size effects originating from background charge compensation have been extensively discussed since the late 1980s in the context of charged defect calculations under PBC.[28,29] In particular, for formation-energy calculations of point defects such as charged vacancies, it is well established that interactions between the localized defect charge and the compensating background charge in finite supercells introduce systematic errors, and numerous correction schemes have been proposed to address this issue.[30–32] These considerations share an essential physical foundation with XANES calculations involving core-hole states, insofar as both involve localized positive charges treated within PBC. In this work, we extend the conceptual framework developed for finite-size effects in charged defect calculations to the analysis of core-hole effects in XANES simulations under PBC.

Previous studies have compared transition energies among different core-hole treatments using atomic-orbital-based methods and PBC-DFT frameworks.[24,33] However, most of these works have focused primarily on relative peak positions within a single molecule, and systematic investigations of transition-energy differences between distinct molecules or absorption sites, namely chemical shifts, remain limited, particularly when accounting for computational settings under PBC. This limitation is especially relevant for the increasingly common approach in which molecules are placed in large supercells and treated using PBC-DFT. In this context, quantitative assessment of how cell size and molecular size affect energy shifts, together with the establishment of reusable benchmarks for future large-scale calculations and data-driven analyses, is of critical importance.

In this study, we therefore systematically examine how different treatments of the core hole, charge compensation schemes, and supercell size influence the calculated absolute transition energies, chemical shifts, and spectral line shapes in molecular XANES calculations under PBC. We first consider an isolated ethane molecule as a model system and quantify the relationship between finite-size effects and spectral shifts by systematically varying the supercell size. Next, under a fixed cell size, we investigate an *n*-alkane series to evaluate how increasing molecular size, specifically the number of carbon atoms, affects transition energies at local absorption sites such as terminal carbons. Finally, we extend our analysis to a set of small molecules and discuss the reproducibility and limitations of chemical shifts between different chemical environments. The insights obtained in this work are expected to provide practical guidelines for selecting computational conditions when applying PBC-DFT-based XANES calculations to molecular, adsorption, and interfacial systems. Furthermore, they offer a methodological foundation for energy alignment in large-scale molecular calculations and the construction of spectral databases.

**Computational Methods**
**2.1 Geometry optimization**
All molecular structures considered in this study were optimized using the quantum chemistry software Gaussian16.[34] Geometry optimizations were performed within density functional theory (DFT), employing the B3LYP exchange–correlation functional[35] together with the 6-31+G(d,p) basis set,[36] which is known to yield reliable equilibrium geometries under these conditions[37]. The default convergence criteria implemented in Gaussian were adopted. The optimized geometries obtained from Gaussian were subsequently used as input structures for the XANES calculations. In the XANES simulations, molecules were placed in large supercells under periodic boundary conditions (PBC) to approximate isolated systems. In subsequent PBC-DFT calculations, the Gaussian-optimized geometries were kept fixed and only the electronic structure was converged self-consistently.

**2.2 Electronic structure calculations under periodic boundary conditions (PBC-DFT settings)**

Electronic structures for XANES calculations were obtained using plane-wave pseudopotential DFT. Larger supercells were systematically tested to confirm the convergence of transition energies with respect to cell size. In this study, supercell sizes ranging from approximately 8 to 30 Å were employed. These cell dimensions ensure a minimum intermolecular separation exceeding 6–15 Å, thereby substantially reducing artificial electrostatic interactions between periodic images in the case of neutral systems.

It is important to emphasize that the finite-size dependence discussed in this work originates from long-range Coulomb interactions under periodic boundary conditions rather than from the detailed radial description of the core wavefunction. To investigate this effect systematically, sufficiently large supercells with extended vacuum regions up to 30 Å were employed for molecules of different sizes, and the transition energies were carefully analyzed as a function of cell dimension. Within the ΔSCF framework adopted here, transition energies are determined by total-energy differences between well-defined initial and final states. Therefore, the essential physics governing the supercell-size scaling is independent of whether a pseudopotential or an all-electron formalism is used. Base on these considerations, All calculations were performed with CASTEP,[14] adopting the PBE exchange–correlation functional.[38] Ultrasoft pseudopotentials[39] were used, with a plane-wave energy cutoff of 500 eV. Brillouin-zone sampling was restricted to the $\Gamma$ point.

**2.3 Core-hole treatments**

To account for core-hole effects associated with core-level excitation in XANES, multiple final-state core-hole models were compared. Specifically, we employed (i) the full core-hole (FCH) and (ii) the excited core-hole (XCH) approaches. Schematic illustrations of the FCH and XCH models investigated in this work are shown in Fig. 1.

In the FCH approach, the final state is constructed by removing one core electron from the absorbing atom, and self-consistent calculations are performed using a core-hole pseudopotential. Under periodic boundary conditions, the system becomes positively charged, which leads to divergence of the Coulomb energy. To stabilize the calculations, a uniform compensating background charge (jellium) was introduced to maintain overall charge neutrality. Because this background charge interacts with the localized positive charge associated with the core hole and can introduce supercell-size-dependent finite-size effects in the transition energies, we systematically varied the cell size to quantitatively evaluate this influence.

In the XCH approach, in addition to creating the core hole, the excited electron is retained within the system and placed in the lowest unoccupied state at the bottom of the conduction band, thereby maintaining global charge neutrality. Since no compensating background charge is required, this method potentially avoids the background-charge-induced finite-size effects that may arise in FCH calculations.

Core holes were introduced at the selected absorption sites for each molecule. For the *n*-alkane series, terminal carbon atoms were chosen as absorption sites, and changes in transition energies and spectral line shapes were analyzed as a function of increasing carbon number.

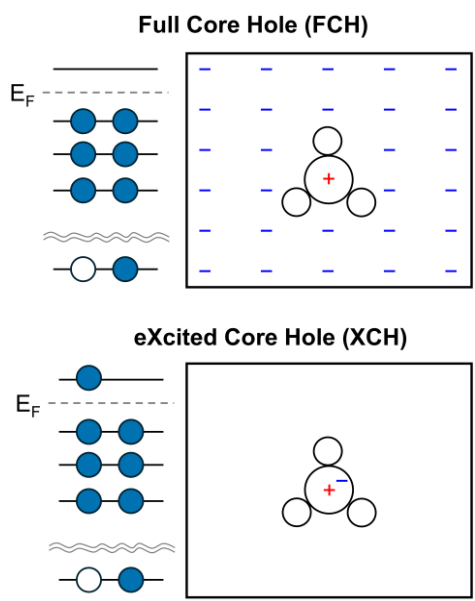

**Figure 1. Schematic of core-hole treatments used for XANES calculations. FCH removes one core electron, yielding a charged final state that is neutralized by a uniform jellium background (blue "−"). XCH retains the excited electron in the lowest unoccupied state, giving a neutral final state without background charge (blue "−" denotes the promoted electron).**

**2.4 XANES spectrum calculation**

XANES spectra were calculated from transition probabilities within the electric dipole approximation. Specifically, transition matrix elements from the core level to unoccupied states were computed, and discrete transitions were placed along the energy axis to obtain absorption intensities. To facilitate comparison of spectral line shapes, the spectra were convoluted with a Gaussian function to mimic core-hole lifetime broadening and experimental energy resolution. A broadening width of 0.4 eV was employed. Polarization dependence was treated by isotropic averaging.

**2.5 Transition energies**

Within the pseudopotential framework, core-electron contributions are not explicitly included in the total energy. Consequently, unlike all-electron calculations, transition energies cannot be directly obtained solely from total energy differences between the ground and excited states. Following our previous work,[19] we therefore introduce a correction term accounting for the core contribution in addition to the valence-electron contribution, and define the transition energy $E_{\mathrm{TE}}$ as

$$E_{\text{TE}} = E_{\text{valence}} + E_{\text{core(atom)}}, \qquad (1)$$

where $E_{\text{valence}}$ represents the valence-electron contribution obtained from pseudopotential calculations,

$$E_{\text{valence}} = E_{\text{FS}}^{\text{PP}} - E_{\text{GS}}^{\text{PP}}. \qquad (2)$$

Here, $E_{\text{GS}}^{\text{PP}}$ is the total energy of the ground state calculated using pseudopotentials, and $E_{\text{FS}}^{\text{PP}}$ is the corresponding total energy of the final state with a core hole. For FCH, a charged final state (neutralized by a compensating background charge) was used, whereas for XCH, a neutral final state with the excited electron occupying the lowest unoccupied state was adopted.

The core-electron contribution $E_{\text{core(atom)}}$ was evaluated from isolated-atom calculations performed during pseudopotential generation and defined as

$$E_{\text{core(atom)}} = E_{\text{All orbitals(atom)}} - E_{\text{valence(atom)}}. \qquad (3)$$

where $E_{\text{All orbitals(atom)}}$ is the excitation energy obtained from all-electron atomic calculations, and $E_{\text{valence(atom)}}$ is the corresponding valence contribution from the same atomic calculations. The subtraction in Eq. (3) avoids double counting of the valence contribution already included in $E_{\text{valence}}$ in Eq. (1). Because $E_{\text{core(atom)}}$ is derived from isolated-atom calculations, chemical shifts across different systems are primarily determined by $E_{\text{valence}}$. Thus, $E_{\text{core(atom)}}$ mainly provides an element-specific constant offset, whereas variations in transition energies among different chemical environments are governed by the valence-electron contribution. Using this approach, relative transition energies (chemical shifts) between spectra can be quantitatively reproduced.

To evaluate supercell-size dependence, isolated ethane was employed as a model system, and calculations were performed using cubic cells with side lengths of 8 Å, 10 Å, …, up to 30 Å. For assessing molecular-size dependence, $n$-alkanes ($n$ = 2, 3, …, 10, 15, and 20) were calculated in a fixed rectangular supercell of 37 Å × 12 Å × 20 Å, and transition energies and spectra at terminal carbon sites were compared.

**Results and Discussion**

**3.1 Supercell-size dependence for ethane**

To quantify finite-size effects arising in isolated-molecule calculations under periodic boundary conditions, we placed an ethane molecule in a cubic supercell and calculated C K-edge XANES spectra while systematically varying the cell edge length $L$ from 8 Å to 30 Å in increments of 2 Å. Figure 2 shows the spectra obtained using the FCH and XCH approaches. For ease of comparison, the position of the first main peak in each spectrum is marked by triangles, and the corresponding peak position at $L = 30$ Å is indicated by a dotted line.

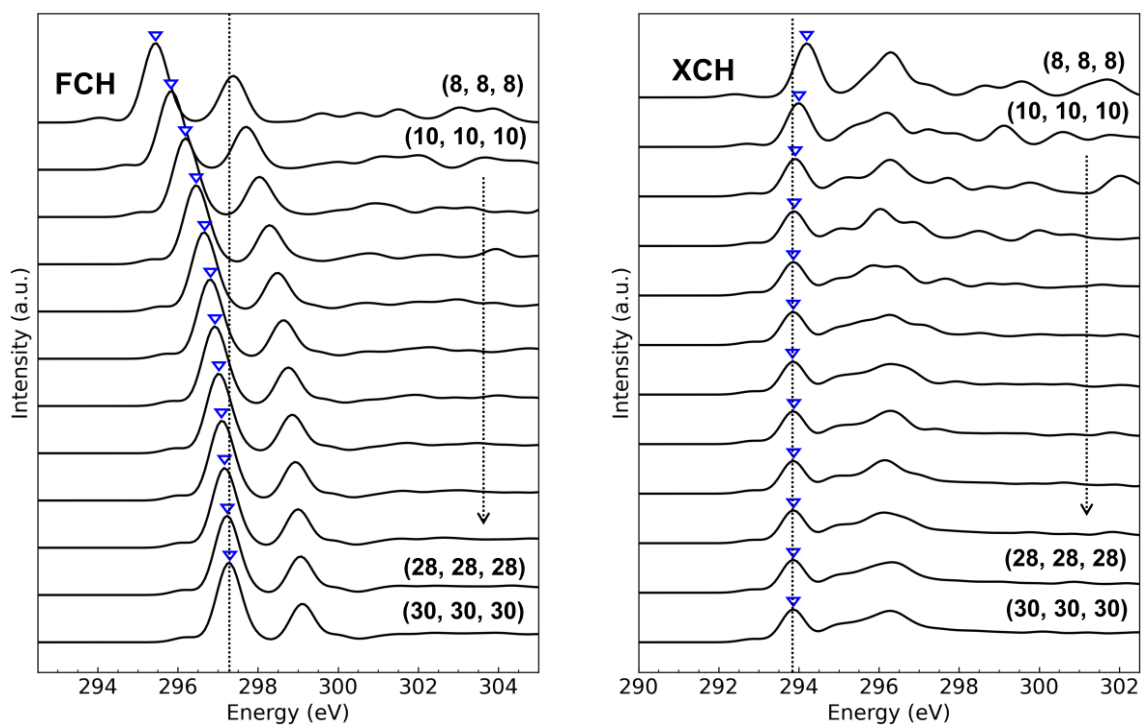

**Figure 2.** C K-edge XANES spectra of ethane calculated in cubic supercells with edge length $L = 8$–$30$ Å (2 Å step). Spectra obtained with FCH (left) and XCH (right) are shown. The first main-peak positions are marked by blue triangles, and the dotted line indicates the first-peak energy at $L = 30$Å.

As clearly seen in Fig. 2, the FCH results exhibit a systematic shift of the first main peak toward higher energies as the cell size increases, and even at $L = 30$ Å the peak position has not yet fully converged. In contrast, for XCH, although a small energy variation is observed for the smallest supercells, the peak position becomes essentially constant for $L \gtrsim 14$ Å, indicating practical convergence with respect to cell size. The relatively large shift observed for XCH in the smallest cell is attributed to insufficient separation to suppress spurious self-interaction of the core hole under periodic boundary conditions, suggesting that an 8 Å cell is inadequate for this purpose (as discussed in more detail below). This contrast originates from charge treatment: FCH employs a charged final state neutralized by a jellium background, whereas XCH constructs a neutral final state by retaining the excited electron, thereby reducing long-range Coulomb artifacts under PBC.

To quantitatively analyze the energy shifts observed in Fig. 2, we plotted the energy of the first main peak as a function of $1/L$, where $L$ denotes the supercell edge length (Fig. 3a). The XCH results remain nearly flat with respect to $1/L$, indicating that the peak position is essentially insensitive to increasing cell size. In contrast, the FCH results exhibit a clear linear dependence, with the first main peak energy varying approximately proportionally to $1/L$. This behavior is consistent with the well-known finite-size effects that arise when charged systems are treated under periodic boundary

conditions, which primarily manifest as terms proportional to $1/L$.[29]

We further examined the $1/L$ dependence of the transition energy $E_{\mathrm{TE}}$, defined as the total energy difference between the ground and final states used in the spectral calculations (Fig. 3b). Similar to the first main peak, $E_{\mathrm{TE}}$ remains nearly constant for XCH, whereas FCH exhibits a pronounced linear dependence on $1/L$. These results indicate that the supercell-size dependence of the XANES energy axis in this system is predominantly governed by the cell-size dependence of $E_{\mathrm{TE}}$.

It should be noted, however, that for small supercells ($L \lesssim 14$ Å), the variations in the first main peak energy do not simply track those of the transition energy. In this regime, the electronic structure of the molecule, particularly the unoccupied states, is influenced by interactions with periodic images, leading to changes in the peak positions that cannot be explained solely by shifts in the energy reference. In other words, distortions of the spectral lineshape become non-negligible. In general, to sufficiently suppress core-hole self-interaction under periodic boundary conditions, a core-hole separation of at least ~10 Å is recommended. The convergence behavior observed for XCH in this study, with stabilization occurring at $L \approx 14$ Å or larger, is consistent with this empirical guideline.[19]

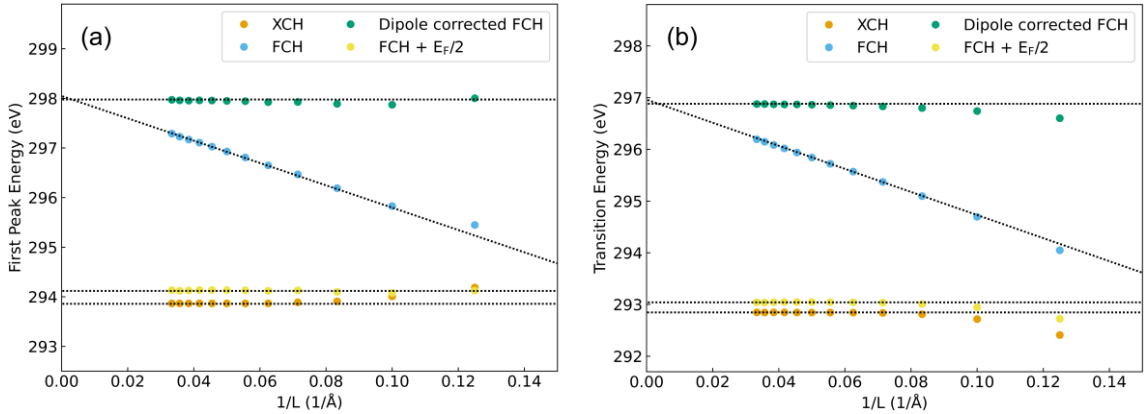

**Figure 3.** Cell-size dependence of (a) the first main-peak energy and (b) the transition energy $E_{\mathrm{TE}}$ for the ethane C K-edge XANES, plotted as a function of $1/L$.

It is well established that treating charged systems under periodic boundary conditions introduces systematic errors in the total energy due to interactions with the compensating jellium background, an issue that has been extensively discussed in the context of point-defect calculations.[28,29] In particular, Makov and Payne demonstrated that, for a charged isolated molecule placed in a cubic supercell, the total energy can be expanded as a sum of a leading term proportional to $1/L$ and higher-order terms (e.g., $1/L^3$) as follows.[29]

$$E = E_0 - \frac{q^2 \alpha}{2L} - \frac{2\pi qQ}{3L^3} + O(L^{-5}) \quad (4)$$

where $E$ is the DFT-calculated total energy, $E_0$ is the true isolated-system energy, $q$ is the net charge (here $+1$ due to a single core hole), $\alpha$ is the Madelung constant (2.8373 for a point charge in a cubic cell[25]), and $Q$ is the quadrupole moment. The nearly linear $1/L$ dependence observed for both the first main peak and the transition energy in FCH strongly suggests that the leading finite-size term dominates in this regime.

Indeed, when the first-order Makov–Payne correction (the $1/L$ term) is applied to the FCH results, the energies become nearly constant, as shown by the green dots in Fig. 3, indicating that most of the supercell-size dependence is removed. This demonstrates that the pronounced cell-size dependence in FCH primarily originates from long-range Coulomb interactions associated with the charged final state neutralized by the background charge. After this correction, the absolute transition energies obtained from XCH and FCH still differ by approximately 4 eV, which can be attributed to the work-function-related energy associated with removing an electron in the FCH scheme.

Having established that the dominant finite-size effect in FCH arises from the leading $1/L$ term and can largely be eliminated by a Makov–Payne-type first-order correction, we next evaluate the contribution of higher-order terms, in particular the quadrupole term proportional to $1/L^3$ in Eq. (4).

Figure 4 shows the difference in transition energy,

$$\Delta E_{\text{TE}}(L) = E_{\text{TE}}(L) - E_{\text{TE}}(30 \text{ Å}) \tag{5}$$

plotted as a function of $1/L^3$, where $E_{\text{TE}}$ corresponds to the values shown in Fig. 3b. For the dipole-corrected FCH results, $\Delta E_{\text{TE}}$ decreases approximately linearly with increasing $1/L^3$, indicating that after removal of the leading term, the quadrupole contribution becomes dominant. In other words, while the $1/L$ term is the primary source of finite-size effects in FCH, higher-order terms predicted by Eq. (4) remain at an observable level after first-order correction. Nevertheless, the magnitude of this residual rapidly diminishes with increasing cell size, and for $L \gtrsim 15$ Å, $|\Delta E_{\text{TE}}|$ generally falls well below 0.1 eV. Therefore, when adopting practically relevant supercell sizes for isolated-molecule approximations (typically on the order of 15–20 Å), residual contributions arising from the quadrupole term can be considered negligible for most applications.

In contrast, XCH does not exhibit a clear $1/L^3$ dependence in $\Delta E_{\text{TE}}$, and the energy differences rapidly approach zero even for relatively small cell sizes. This behavior can be attributed to the fact that XCH constructs a globally neutral final state, thereby intrinsically suppressing finite-size effects associated with long-range Coulomb interactions characteristic of charged systems.

As discussed above, the pronounced supercell-size dependence observed in FCH originates primarily from long-range Coulomb interactions associated with the charged final state, and can largely be eliminated by applying a Makov–Payne-type first-order correction. However, this procedure requires calculations at multiple cell sizes and therefore incurs increased computational cost in practical applications. To provide a simpler alternative for mitigating supercell-size effects, we introduce an empirical correction in which a constant proportional to the Fermi level of the final state is added to the energy. In this work, we adopt a correction of $E_\text{F}/2$. This choice reflects how the chemical potential of the charged final state enters the total-energy difference when comparing states with different electron numbers, and captures the leading finite-size contribution associated with the Fermi-level shift.

Specifically, when $E_\text{F}/2$ is added to the transition energies and peak positions obtained from FCH calculations, the supercell-size dependence is significantly reduced, as shown by the yellow dots in Fig. 3, yielding convergence behavior comparable to that of XCH. Furthermore, Fig. 4 (yellow dots) shows the corresponding energy differences $\Delta E_\text{TE}(L)$, defined in Eq. (5), indicating that FCH+$E_\text{F}/2$, similar to XCH, rapidly approaches zero even for relatively small cell sizes and does not exhibit a clear residual dependence proportional to $1/L^3$ within the range examined in this study. This suggests that FCH+$E_\text{F}/2$ effectively suppresses higher-order finite-size effects that become apparent after removal of the leading term, in a manner comparable to XCH.

This behavior can be interpreted as follows: in charged calculations employing a compensating jellium background, the reference electrostatic potential (i.e., the energy zero) can shift depending on the supercell conditions, whereas the Fermi-level-based correction effectively cancels this offset. Unlike the Makov–Payne correction, FCH + $E_\text{F}/2$ can be applied directly to results obtained from a single supercell calculation, which is advantageous from a practical standpoint. It therefore represents a potentially useful and computationally efficient approach for mitigating supercell-size effects in periodic-boundary calculations of molecular and cluster systems.

Taken together, for the C K-edge XANES of ethane, we demonstrate that either employing XCH or applying appropriate corrections to FCH (Makov–Payne or $E_\text{F}/2$) can reduce charge-induced supercell-size dependence to sufficiently small levels.

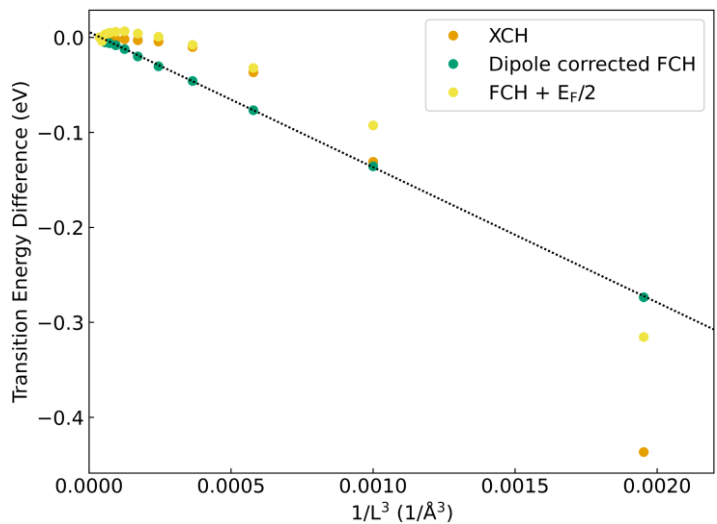

**Figure 4. Residual cell-size dependence of the ethane transition energy, plotted as $\Delta E_\text{TE}(L)$ (Eq.**

(5)) versus $1/L^3$, comparing XCH, dipole-corrected FCH, and FCH + $E_F/2$.

## 3.2 Molecular-size dependence in the *n*-alkane series

Based on the supercell-size analysis using ethane in the previous section, we demonstrated that FCH exhibits pronounced finite-size effects originating from the charged final state, leading to systematic variations in transition energies that follow approximately $1/L$ (and $1/L^3$ after correction), whereas XCH converges rapidly even for relatively small supercells (Figs. 2–4). Building on these findings, we next investigate how the calculated energy shifts behave when the molecular size is varied while keeping the supercell size fixed.

Specifically, we employed a rectangular supercell large enough to ensure a minimum intermolecular separation exceeding 10 Å even for the largest molecule, and calculated a series of *n*-alkanes (from ethane to *n*-eicosane) under identical conditions. Terminal carbon atoms were chosen as the absorption sites, and both the first main peak energies and transition energies were analyzed (Fig. 5). The calculated spectra for all molecules are shown in Fig. S1. As shown in Fig. S1, in the small-carbon-number regime (*n* = 2–4), the first main peak systematically shifts toward lower energies for both FCH and XCH. This behavior can be attributed to intrinsic changes in the electronic structure associated with molecular elongation, such as modifications in the character and localization of unoccupied states (e.g., σ* orbitals), which alter the energy distribution of final states accessed from the terminal carbon. Because this peak shift appears consistently in both FCH and XCH, it is primarily governed by molecular-size-induced electronic structure changes rather than finite-size effects specific to charged final states.

In contrast, for $n \gtrsim 6$, a clear divergence between FCH and XCH emerges. As shown in Fig. 5a, the first main peak energies obtained from XCH and FCH+$E_F/2$ converge to nearly constant values as the carbon number increases, whereas in FCH the peak continues to shift gradually toward lower energies with increasing molecular size. This discrepancy indicates that the peak-position evolution in FCH cannot be explained solely by intrinsic molecular electronic structure effects. Instead, the residual systematic shift observed in FCH is attributed to finite-size effects associated with the charged calculations, as discussed in the previous section, which re-emerge as the molecular size increases.

This interpretation is consistent with previously reported experimental observations. In a systematic comparison of C K-edge XANES spectra from ethane to hexane (including propane, butane, and pentane), it was shown that the first peak position in ethane is relatively shifted, whereas with increasing carbon number the peak position gradually converges and becomes nearly constant for larger alkanes.[40] In the present study, both XCH and FCH + $E_F/2$ exhibit rapid convergence of the peak position toward a constant value as the carbon number increases, naturally reproducing this experimentally observed trend of "convergence with increasing chain length."

This behavior becomes even clearer in the analysis of the transition energy $E_{TE}$ (Fig. 5b). For XCH

and FCH + $E_F/2$, $E_{TE}$ remains nearly constant with respect to carbon number, indicating that the total energy difference between the final and ground states is evaluated in a stable manner even as molecular size increases. In contrast, FCH exhibits a monotonic decrease in $E_{TE}$ with increasing carbon number. Although the supercell dimensions are fixed, this trend arises because larger molecules modify the effective charge distribution within the cell, including the interaction between the localized core hole and the compensating background charge. As a result, finite-size effects effectively become more pronounced even within the same supercell. In this sense, the molecular-size dependence observed in FCH reflects the same underlying physics discussed in Fig. 4, namely finite-size effects associated with charged final states, albeit manifested under a different computational setting than direct supercell-size variation.

Taken together, these results demonstrate that, for the *n*-alkane series, the peak positions at terminal carbon sites are governed by two distinct contributions: (i) in the small-molecule regime, intrinsic electronic-structure changes associated with molecular elongation dominate, whereas (ii) for larger molecules, systematic energy shifts arising from charged-final-state finite-size effects can persist in FCH. Consequently, when the goal is to compare or align spectra across multiple molecules or absorption sites on a common energy scale, employing XCH (or FCH with appropriate corrections) is advantageous, as it provides stable energy alignment that is largely independent of molecular size.

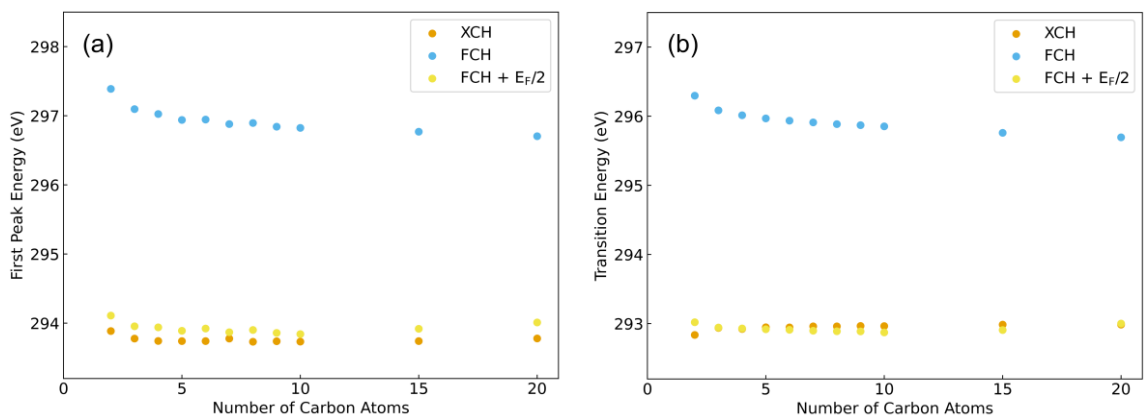

**Figure 5. Carbon-number dependence of (a) the C K-edge first main-peak energy and (b) the transition energy $E_{TE}$ for terminal carbon atoms in *n*-alkanes (ethane to *n*-eicosane), calculated using a fixed supercell. Results obtained with FCH, XCH, and FCH + $E_F/2$ are compared.**

### 3.3 Chemical-shift reproducibility among small molecules

In the previous sections, we demonstrated that energy-axis shifts in FCH are largely governed by finite-size effects associated with charged final states, primarily through the leading $1/L$ term and, after correction, the $1/L^3$ term, whereas such effects are strongly suppressed in XCH (Figs. 2–4). We also showed that, even when the supercell size is fixed and the molecular size is varied, FCH can exhibit systematic variations in transition energies, while XCH and $\mathrm{FCH} + E_\mathrm{F}/2$ display significantly more stable behavior (Fig. 5). These results suggest that the reproducibility of relative transition-energy differences, namely chemical shifts, which is essential for aligning spectra across multiple molecules or absorption sites, strongly depends on the chosen core-hole treatment. In this section, we therefore examine how accurately the relative peak positions arising from chemical differences between molecules can be reproduced.

Figure 6a shows the C K-edge XANES spectra of acetylene, ethylene, benzene, ethane, and n-butane. From top to bottom, the panels correspond to XCH, FCH, $\mathrm{FCH} + E_\mathrm{F}/2$, and experimental spectra. All molecules were calculated using the same supercell size and shape, ensuring that differences originating from cell geometry do not contaminate the intermolecular comparison. Under these conditions, the leading finite-size term proportional to $1/L$ discussed for FCH acts primarily as a common offset across molecules, since the cell size is identical. Moreover, because the molecular sizes are comparable and all systems are treated in sufficiently large supercells, residual contributions arising from the quadrupole term ($\propto 1/L^3$) observed in the previous section are expected to be negligible for intermolecular comparisons. Consequently, this section focuses primarily on the impact of the core-hole approximation itself on chemical-shift reproducibility.

To quantitatively assess this effect, the energies of the first main peak for each molecule were plotted with experimental values on the horizontal axis and calculated values on the vertical axis (Fig. 6b). For direct comparison with experiment, a reference line with unit slope allowing for a constant offset is shown as a dotted line. The first main peaks obtained using FCH deviate substantially from this reference line, indicating that the relative relationships between molecules, namely the chemical shifts, are not adequately reproduced. This behavior reflects the fact that FCH treats a charged final state neutralized by a compensating background charge, causing the energy reference of the final state (i.e., the electrostatic potential zero) to depend sensitively on the molecular electron density distribution and screening. As a result, systematic errors can emerge in intermolecular comparisons.

In contrast, the first main peaks obtained from XCH lie closely along the reference line, indicating that relative peak-position differences between molecules are reproduced with high accuracy. Furthermore, the results obtained using $\mathrm{FCH} + E_\mathrm{F}/2$ agree well with the XCH peak positions and cluster near the reference line. This demonstrates that, even within the FCH framework, introducing a Fermi-level-based correction effectively compensates for the energy-reference offset inherent to charged calculations, enabling chemical shifts to be reproduced with accuracy comparable to XCH. Taken together, these results indicate that XCH provides the most robust description for discussing relative transition-energy differences between molecules, while for FCH, corrections such as $E_\mathrm{F}/2$ are essential to achieve comparable performance.

To confirm that this trend is not limited to the C K-edge, we performed an analogous analysis for the N K-edge (Fig. S2). Similar to the C K-edge results, FCH exhibits substantial deviations from experimental relative trends, whereas XCH and FCH + $E_F/2$ cluster near the reference line, demonstrating significantly improved chemical-shift reproducibility. These findings indicate that the conclusion of this study, namely that stable reproduction of intermolecular relative peak positions can be achieved using either neutral final states (XCH) or FCH supplemented with appropriate energy-reference corrections, holds for both C and N K-edges.

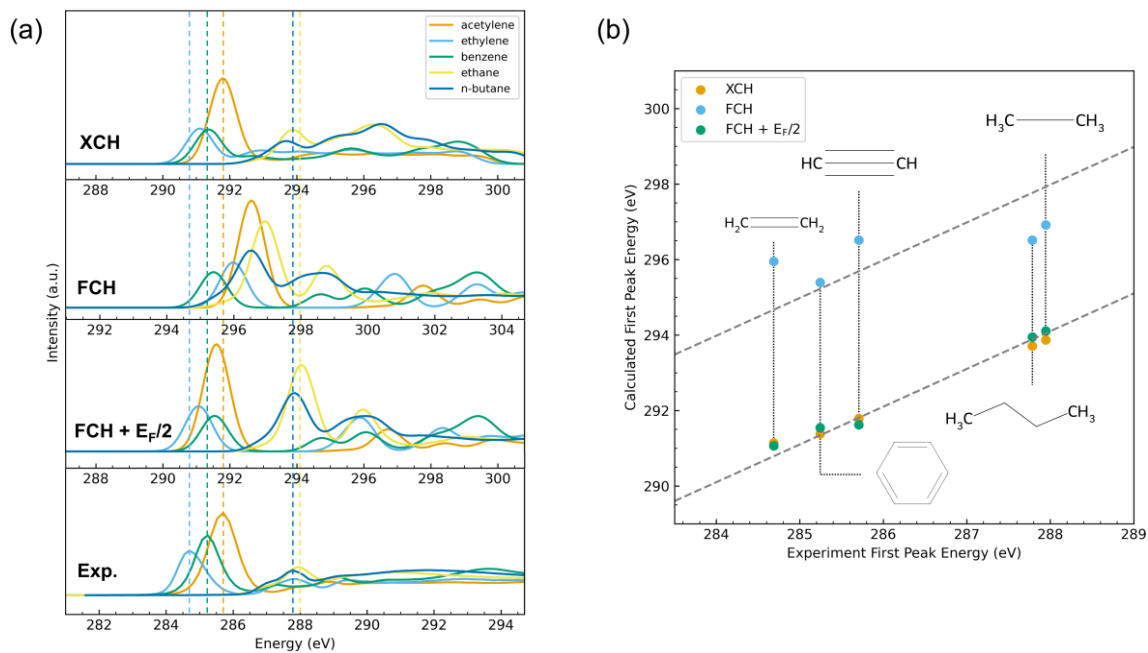

Figure 6. Comparison of calculated and experimental C K-edge XANES for selected small molecules. (a) Calculated spectra obtained with XCH, FCH, and FCH + $E_F/2$, together with the experimental spectra. (b) Correlation plot of the first main-peak energy, with experimental values on the x-axis and calculated values on the y-axis (dotted lines indicate a slope of 1 with an arbitrary offset).

**Conclusions**

We investigated how core-hole treatments (FCH and XCH) and charge-compensation schemes affect $E_{TE}$ and energy alignment in PBC-DFT-based molecular XANES, focusing on supercell-size effects, molecular-size effects, and intermolecular chemical-shift reproducibility.

First, from the supercell-size analysis using ethane, we showed that in FCH the first main peak exhibits systematic shifts with increasing cell size and does not fully converge even at approximately 30 Å, whereas in XCH the peak position becomes essentially constant for $L \approx 14\,\text{Å}$ or larger, indicating practical convergence of finite-size effects. This contrast arises because FCH treats a charged final state neutralized by a jellium background, while XCH constructs a neutral final state and is therefore less susceptible to long-range Coulomb interactions. We further demonstrated that the pronounced cell-size dependence in FCH is dominated by the leading finite-size term proportional to $1/L$, which can largely be removed by a Makov–Payne-type correction. After removing the leading term, the residual $1/L^3$ contribution becomes visible but is below 0.1 eV for $L \geq 15$Å. In addition, we introduced a simple energy-reference correction based on the final-state Fermi level, $E_\text{F}/2$, and confirmed that this approach significantly mitigates the cell-size dependence in FCH. Unlike the Makov–Payne correction, which requires calculations at multiple cell sizes, the $E_\text{F}/2$ correction can be directly applied to results obtained from a single supercell, making it particularly attractive for high-throughput calculations and large-scale systems.

Second, in the *n*-alkane series with a fixed supercell size, we found that in the small-molecule regime, peak-position variations primarily reflect intrinsic electronic-structure changes associated with molecular elongation. In contrast, for larger molecules, systematic energy shifts persist only in FCH. In particular, while $E_\text{TE}$ remains nearly constant with respect to carbon number in XCH and FCH+$E_\text{F}/2$, it decreases monotonically in FCH as molecular size increases. This behavior indicates that finite-size effects associated with the charged final state can re-emerge even within a fixed supercell as the molecular size changes. Consequently, for applications requiring alignment and comparison of spectra across multiple molecules or absorption sites on a common energy scale, XCH or appropriately corrected FCH provides a clear advantage.

Third, in assessing intermolecular chemical-shift reproducibility, we showed that FCH fails to reproduce the relative experimental peak positions, whereas XCH accurately captures these relationships, and FCH+$E_\text{F}/2$ closely matches the XCH results. This demonstrates that introducing an energy-reference correction enables FCH to achieve chemical-shift accuracy comparable to XCH. Similar trends were observed for both the C K-edge and N K-edge, indicating that these conclusions hold at least for these two edges.

Taken together, our results provide practical guidelines for molecular XANES calculations based on PBC-DFT when quantitative energy alignment and intermolecular chemical shifts are of interest: (i) preferentially employ XCH, which naturally yields a neutral final state; (ii) when using FCH, combine finite-size corrections (at least the leading-order term) with an energy-reference correction such as $E_\text{F}/2$; and (iii) ensure sufficiently large supercells to separate core holes adequately (in this study, $L \gtrsim 15$Å yields residual errors below 0.1 eV) and explicitly assess any remaining finite-size contributions. These insights are expected to serve as a practical foundation for selecting computational conditions in PBC-DFT-based XANES studies of molecular, adsorption, and interfacial systems, as well as for establishing robust energy-alignment procedures in large-scale molecular calculations and spectral database construction.

**Acknowledgments**

This study was supported by the Ministry of Education, Culture, Sports, Science and Technology (MEXT). Prof. Kiyou Shibata (Nagoya University) would be acknowledged for helpful discussions.

**Conflict of Interest**

The authors have no conflicts to disclose.

**Author Contributions**

**Yu Fujikata**: Data curation (lead); Formal analysis (lead); Investigation (lead); Methodology (lead); Software (lead); Validation (lead); Visualization (lead); Writing – original draft (lead); Writing – review & editing (equal). **Yasuji Muramatsu**: Conceptualization (lead); Methodology (supporting); Writing – review & editing (equal). **Teruyasu Mizoguchi**: Supervision (lead); Conceptualization (supporting); Methodology (supporting); Writing – review & editing (equal).

**Data Availability**

The data that support the findings of this study are available from the corresponding author upon reasonable request.

# Supplementary Information

Finite-size effects and energy alignment in molecular XANES under periodic boundary conditions: A systematic comparison of core-hole treatments

Yu Fujikata[1,2,a)], Yasuji Muramatsu[3] and Teruyasu Mizoguchi[1,a)]

[1]Institute of Industrial Science, The University of Tokyo, Tokyo 153-8505, Japan.
[2]Science & Innovation Center, Mitsubishi Chemical Corporation, Yokohama 227-8502, Japan.
[3]Graduate School of Engineering and Laboratory of Advanced Science and Technology for Industry (LASTI), University of Hyogo 2146 Shosha, Himeji 671–2201, Japan.

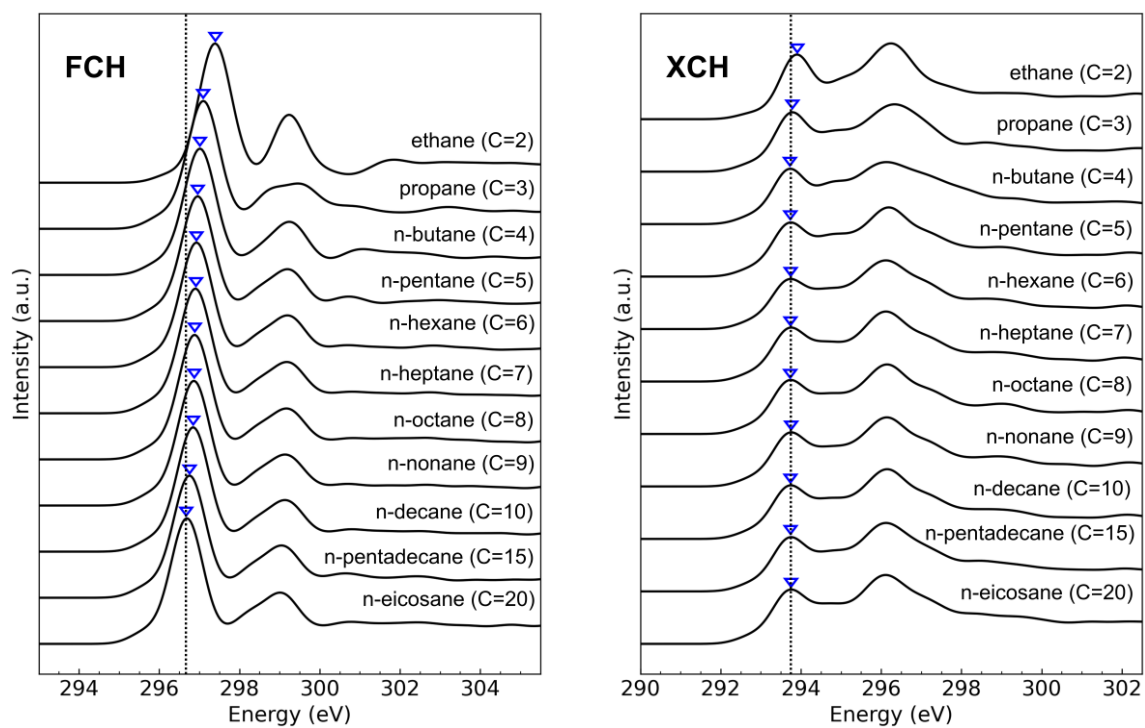

**Figure S1.** Calculated C K-edge XANES spectra of edge carbon atoms for n-alkanes from ethane to n-eicosane obtained using (left) the full core-hole (FCH) and (right) the excited core-hole (XCH) approaches. Blue triangles mark the positions of the first main peak for each spectrum, and the dotted vertical line indicates the first-peak position of ethane for reference.

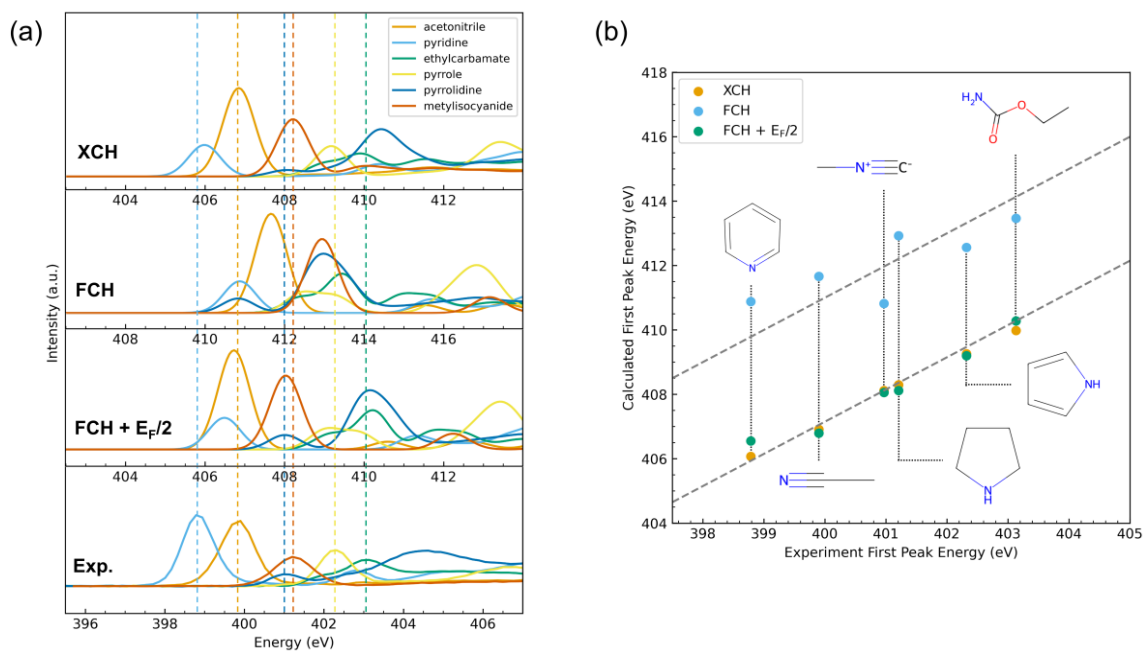

**Figure S2.** Comparison of calculated and experimental N K-edge XANES for selected small molecules. (a) calculated spectra obtained with XCH, FCH, and FCH + $E_F/2$, together with the experimental spectra. (b) correlation plot of the first main-peak energy, with experimental values on the x-axis and calculated values on the y-axis (dotted lines indicate a slope of 1 with an arbitrary offset).